\documentclass[aps,prl,superscriptaddress,twocolumn]{revtex4}
\usepackage{graphicx}
\usepackage{amsmath}
\usepackage{amssymb}
\usepackage{booktabs}	
\usepackage{times}

\def\be{\begin{equation}}
\def\ee{\end{equation}}

\begin{document}

\title{Molecular lattice clock with long vibrational coherence}

\author{S. S. Kondov}
\affiliation{Department of Physics, Columbia University, 538 West 120th Street, New York, NY 10027-5255, USA}
\author{C.-H. Lee}
\affiliation{Department of Physics, Columbia University, 538 West 120th Street, New York, NY 10027-5255, USA}
\author{K. H. Leung}
\affiliation{Department of Physics, Columbia University, 538 West 120th Street, New York, NY 10027-5255, USA}
\author{C. Liedl}
\altaffiliation[Present address:  ]{Department of Physics, Humboldt University of Berlin, Newtonstra\ss e 15, 12489 Berlin, Germany}
\affiliation{Department of Physics, Columbia University, 538 West 120th Street, New York, NY 10027-5255, USA}
\author{I. Majewska}
\affiliation{Quantum Chemistry Laboratory, Department of Chemistry, University of Warsaw, Pasteura 1, 02-093 Warsaw, Poland}
\author{R. Moszynski}
\affiliation{Quantum Chemistry Laboratory, Department of Chemistry, University of Warsaw, Pasteura 1, 02-093 Warsaw, Poland}
\author{T. Zelevinsky}
\email{tanya.zelevinsky@columbia.edu}
\affiliation{Department of Physics, Columbia University, 538 West 120th Street, New York, NY 10027-5255, USA}

\begin{abstract}     
Atomic lattice clocks have spurred numerous ideas for tests of fundamental physics, detection of general relativistic effects, and studies of interacting many-body systems.  On the other hand, molecular structure and dynamics offer rich energy scales that are at the heart of new protocols in precision measurement and quantum information science.  Here we demonstrate a fundamentally distinct type of lattice clock that is based on vibrations in diatomic molecules, and present coherent Rabi oscillations between weakly and deeply bound molecules that persist for 10's of milliseconds.  This control is made possible by a state-insensitive magic lattice trap that weakly couples to molecular vibronic resonances and enhances the coherence time between molecules and light by several orders of magnitude.  The achieved quality factor $Q=8\times10^{11}$ results from 30-Hz narrow resonances for a 25-THz clock transition in Sr$_2$.  Our technique of extended coherent manipulation is applicable to long-term storage of quantum information in qubits based on ultracold polar molecules, while the vibrational clock enables precise probes of interatomic forces, tests of Newtonian gravitation at ultrashort range, and model-independent searches for electron-to-proton mass ratio variations.
\end{abstract}
\date{\today}
\maketitle

Progress in molecular quantum state control \cite{YeMosesNPhys17_PolarMoleculeFrontiers} has recently led to molecular laser cooling \cite{DeMilleBarryNature14_SrFMOT}, novel approaches to precision measurement \cite{CornellCairncrossPRL17_JILAeEDMI}, as well as studies of ultracold chemistry phenomena \cite{YeOspelkausScience10_KRbReactions,ZelevinskyMcDonaldNature16_Sr2PD}, many-body physics \cite{YeYanNature13_KRbLatticeSpinModel}, and quantum information \cite{ZwierleinParkScience17_NaKNuclearSpinCoherence}.  Here we expand this control by extending the molecule-light vibrational coherence time in an optical lattice by over a thousandfold through a general and widely applicable technique, and utilizing this enhanced coherence to demonstrate a vibrational molecular clock.

Atomic clocks have proven to be extraordinarily precise scientific measurement tools \cite{YeCampbellScience17_3DClock,LudlowSchioppoNPhot17_DualYbClock,KatoriNemitzNPhot16_YbSrClockRatio}, enabling measurements that address fundamental constants and dark energy \cite{RosenbandScience08,BlattPRL08,GillGodunPRL14_YbIonClockConstantVariations,PeikHuntemannPRL14_YbIonCsClockConstantVariations}, general relativity \cite{WinelandScience10_Relativity}, gravitational waves \cite{YeKolkowitzPRD16_GravityWavesLatticeClocks}, and many-body physics \cite{YeKolkowitzNature17_SrClockSOC}.  Some of the experiments depend on the physics of the clock mechanism.  For example, clocks based on electronic transitions can constrain the stability of the fine structure constant while those based on hyperfine transitions help measure the stability of the electron-to-proton mass ratio.  Molecules possess a more extensive set of internal degrees of freedom than atoms including vibrations and rotations.  A clock based on molecular vibrations can access fundamental measurements that are out of reach for atomic clocks \cite{TakahashiBorkowskiPRA17_YbPABeyondBO,BorkowskiPRL18,ZelevinskyPRL08,ChardonnetShelkovnikovPRL08_muStability,KorobovSchillerPRL14_H2IonClocks,OdomStollenwerkAtoms18_TeHIonOpticalPumping,HannekePRA16_O2Ion}.
These include searches for new forces \cite{UbachsSalumbidesPRD13_FifthForcesMolecPrecisMeasts}, model-independent tests of the electron-to-proton mass ratio stability \cite{SchwerdtfegerBeloyPRA11_AlphaVarInSr2}, and tests of quantum electrodynamics in bound systems \cite{UbachsSalumbidesPRL11_H2MoleculeQED,SalumbidesTrivikramPRL18_T2QED}.  Many of the same features that enhance molecular clocks are vital for long-term storage of quantum information in molecules \cite{ZwierleinParkScience17_NaKNuclearSpinCoherence}.

\begin{figure}
\includegraphics*[trim = 0in 2.5in 3.4in 3.8in, clip, width=3.375in]{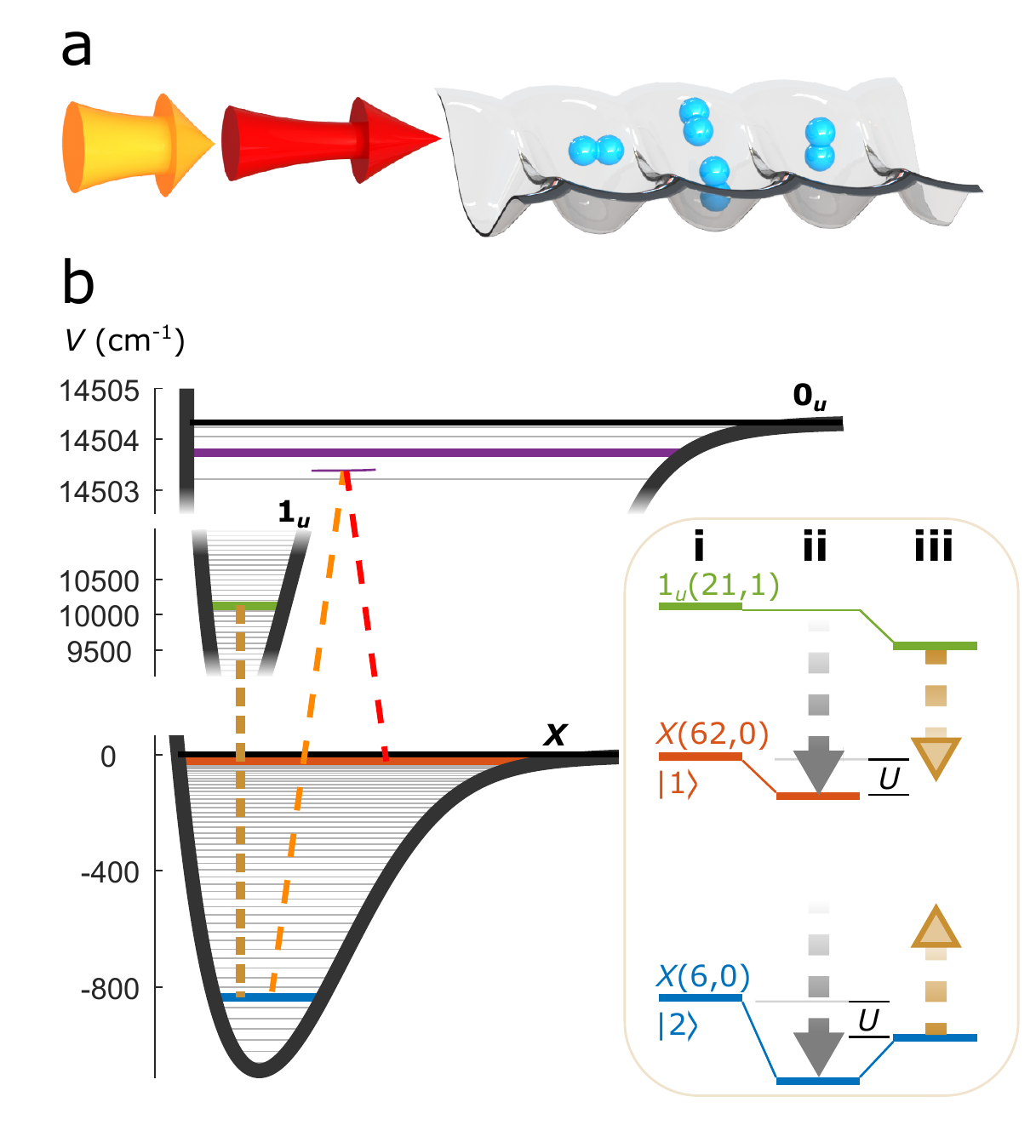}
\caption{Vibrational molecular lattice clock.  a) $^{88}$Sr$_2$ molecules are trapped in a one-dimensional optical lattice and probed in the Lamb-Dicke regime with two-color light.  b) The transition between clock states $|1\rangle$ and $|2\rangle$ is driven by an off-resonant Raman process (dashed red and orange lines).  The lattice (short-dashed yellow line, $\sim920$ nm) couples the lower clock state to one of $1_u$ vibrational levels, ensuring decoherence-free magic trapping.  $^{88}$Sr$_2$ molecular potentials:  ground-state X$0_g^+$ correlates to the $^1S{+^1S}$ atomic threshold and singly-excited $0_u^+$ and $1_u$ correlate to $^1S{+^3P}_1$.  Inset:  clock state energies (i) are shifted down by different amounts in the lattice as indicated with grey arrows (ii), but at a magic wavelength $|2\rangle$ is additionally up-shifted (yellow arrows) to equalize the trap depths of $|1\rangle$ and $|2\rangle$ (iii).}
\label{fig:Fig1}
\end{figure}
The molecular clock is based on vibrational excitations of $^{88}$Sr$_2$ confined in a one-dimensional lattice, as shown in Fig. \ref{fig:Fig1}(a).  Two laser beams copropagate along the lattice axis and probe the clock transition in the anti-Stokes Raman configuration.  Tightly trapping neutral molecules in the lattice (the Lamb-Dicke parameters are below 0.15 for lattice light and 0.012 for probe light) affords a large signal-to-noise ratio while eliminating motional effects that lead to rapid decoherence.

Figure \ref{fig:Fig1}(b) illustrates the details of the clock.  Approximately $7000$ molecules are prepared from laser-cooled $^{88}$Sr atoms at $\sim5$ $\mu$K, predominantly in the electronic ground states X$0_g^+(v=-1, J=\{0,2\})$ \cite{ZelevinskyReinaudiPRL12_Sr2}, where only $J=0$ is used in this work and serves as the upper clock state $|1\rangle$.  The vibrational quantum number is negative if counted down from the threshold; $J$ and $M$ are the total angular momentum and its projection onto the quantization axis that is set by the linear probe light polarization and a weak magnetic field.
The 25.1 THz Raman clock transition is driven by 689.4 and 651.9 nm light, as indicated with dashed red and orange lines, reaching the deeply bound ($v=6$) lower clock state $|2\rangle$.  The 689.4 nm probe light is phase-locked to the narrow-line Sr cooling light with a $\sim3\times10^{-13}$ fractional optical linewidth which also stabilizes the repetition rate of a femtosecond frequency comb.  The 651.9 nm probe light is phase-locked to the comb.  Depending on the chosen Raman wavelengths, Sr$_2$ vibrational clock frequencies can range from $\sim1$ GHz to $\sim30$ THz, spanning the molecular potential depth.
The Raman detuning from the intermediate state $0_u^+(v'=-4,J'=1,M'=0)$ is 25 MHz while the natural width of this state is $\sim20$ kHz.
The natural lifetimes of the clock states are expected to be very long, $\gtrsim10^6$ years, since $(v,J=0)$ decays predominantly to $(v''=v-2,J''=2)$ through rovibrational electric quadrupole transitions, as described in Supplementary Information (SI).  Molecule detection is performed by photodissociating $|1\rangle$ molecules (directly to an excited atomic continuum or by exciting to a self-dissociating molecular state) and absorption imaging the Sr photofragments \cite{ZelevinskyMcGuyerNJP15_Sr2Spectroscopy}.  The duration of a single experiment on a fresh molecular sample is $\sim2$ s.

The key enabling concept for the clock is a molecular state-insensitive (magic) trap.  Magic trapping was a major breakthrough for optical atomic lattice clocks \cite{YeSci08}.  Previously we have demonstrated magic trapping for ultranarrow one-photon molecular transitions \cite{ZelevinskyMcGuyerNPhys15_Sr2M1} which enabled a new regime of molecular metrology.  However, the light-shift cancellation mechanism was similar to that used for atomic clocks, since the molecular clock states were near atomic thresholds with different electronic and spin characters.  Other schemes for state-insensitive trapping of molecules typically involve specific lattice polarizations or intensities \cite{DeMilleKotochigovaPRA10_PolarMoleculesStateInsensTrapping,JinNeyenhuisPRL12_KRbAnisotropicPolarizability,KotochigovaLiPRA17_NaKMagicTrappingWithFields,NiRosenbandOE18_MoleculeMagicTrappingEllipticalPolariz}, parameters that are difficult to control with a very high precision.
For scalar ($J=0$) clock states, primarily the lattice wavelength can be used for light shift tuning.  In contrast to atomic clocks, $|1\rangle$ and $|2\rangle$ belong to the same electronic potential, resulting in nearly parallel nonresonant ac polarizabilities versus wavelength.  Using polarizability crossing points near vibronic molecular resonances was proposed \cite{ZelevinskyPRL08,KajitaPRA11_YbLiMagicTrappingProposal} but not demonstrated prior to this work.

To implement state-insensitive trapping, the lattice (short-dashed yellow line in Fig. \ref{fig:Fig1}(b)) couples $|2\rangle$ to a deeply bound $1_u$ vibrational level at the blue tail of the resonance.  The inset to Fig. \ref{fig:Fig1}(b) shows the unperturbed clock states and the $1_u$ state to which $|2\rangle$ is coupled by the lattice (i).  The clock state light shifts are also shown (ii), along with a reduced light shift of $|2\rangle$ when the lattice is tuned near the vibronic resonance, resulting in equal trap depths for $|1\rangle$ and $|2\rangle$ and, therefore, an unshifted clock resonance that is free of lattice-induced inhomogeneous broadening (iii).

\begin{figure}
\includegraphics*[trim = 0in 3.8in 3.3in 2.5in, clip, width=3.375in]{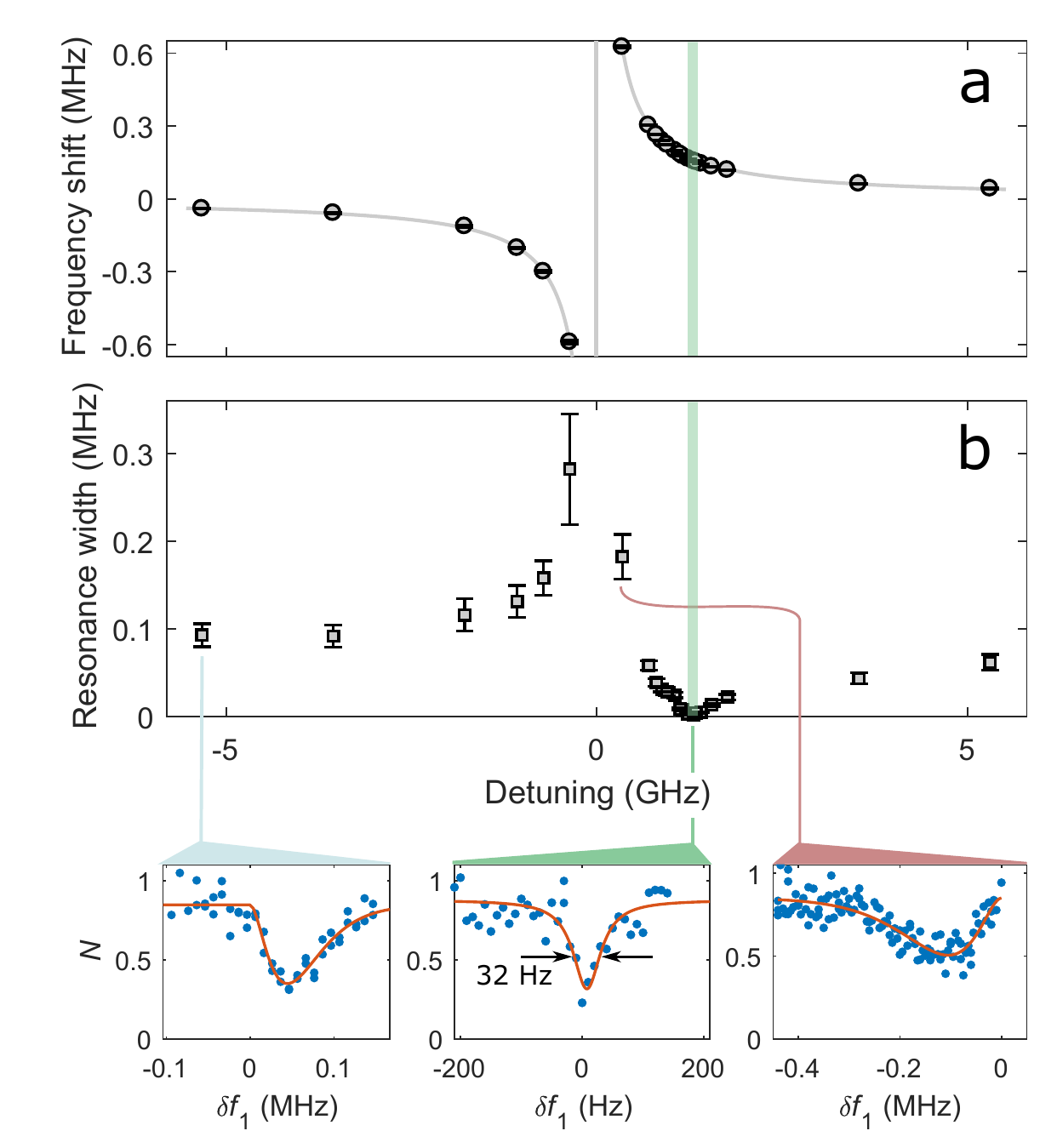}
\caption{Magic lattice for the molecular clock.  a) The measured light shift of clock state $|2\rangle$ versus lattice light detuning shows a polarizability resonance originating from coupling to $1_u(v'=20)$.  The vertical axis refers to the detuning of the 689.4 nm probe (dashed red line in Fig. \ref{fig:Fig1}(b)).  The dispersive fitting function is described in Supplementary Information (SI).  The polarizability of $|2\rangle$ equals that of $|1\rangle$ near the high-density points $\sim1$ GHz above resonance.  b)  Clock transition linewidth drops by over $10^3$ at the magic wavelength.  Three clock-transition spectra are shown versus the probe detuning, each connected to its linewidth.  The magic-wavelength resonance has a fitted Lorentzian width of 32(3) Hz and signal-to-noise ratio of 10, while the resonances at nonmagic wavelengths exhibit thermal broadening of $\sim0.1$ MHz with an asymmetry determined by the relative magnitude and sign of the clock-state trap depths \cite{ZelevinskyMcDonaldPRL15_Sr2LatticeThermometry} as described in SI.  The error bars in panels (a) and (b) correspond to the standard errors of Lorentzian fits to the clock spectra.}
\label{fig:Fig2}
\end{figure}
Four deeply bound states of the $1_u$ potential ($v'=\{19$-$22\},J'=1$) were found by tuning the lattice wavelength between 926 and 907 nm.  The assignment of vibrational labels was based on a comparison of measured transition frequencies and vibrational spacings with \textit{ab initio} calculations \cite{MoszynskiSkomorowskiJCP12_Sr2Dynamics}.  The vibronic resonances, separated by $\sim6$ nm, were detected via the light shift of $|2\rangle$, as plotted in Fig. \ref{fig:Fig2}(a) for the 919.7 nm resonance ($v'=20$) and in SI for the other resonances.  As the lattice frequency is scanned by 10 GHz across this resonance, the width of the clock transition drops by over three orders of magnitude as shown in Fig. \ref{fig:Fig2}(b).  Three sample spectra are presented where the lattice is tuned to the red side of the resonance (left; $|2\rangle$ is more deeply trapped than $|1\rangle$), to the blue side (right; $|2\rangle$ is less deeply trapped), and further to the blue where the clock state polarizabilities are matched (center; $|1\rangle$ and $|2\rangle$ are equally trapped).  The magic-wavelength spectrum, obtained with low probe power corresponding to an effective Rabi frequency $\Omega=12(1)$ Hz, has a linewidth of 32(3) Hz width and a signal-to-noise ratio of 10.  This width yields a quality factor $Q=8\times10^{11}$, hence the fractional clock uncertainty is expected to average down as $2\times10^{-13}/\sqrt{\tau}$ in $\tau$ seconds, reaching 1 Hz in 20 s.  The noise on the clock spectra can be further suppressed by implementing zero-background detection schemes, as in atomic clocks.

Consequences of working near trap-induced resonances include a potentially steep dependence of the differential light shift on lattice frequency.  For the 919.7 nm resonance, the measured light shift is $\sim$1 Hz / 20 kHz which is several orders of magnitude more sensitive to lattice detuning than for atomic lattice clocks.  However, lattice frequency stabilization below the 10 kHz level is not prohibitive.  We achieve the necessary stability by phase-locking the lattice to the clock laser via the frequency comb.

\begin{figure}
\includegraphics*[trim = 0in 2.5in 3in 2in, clip, width=3.375in]{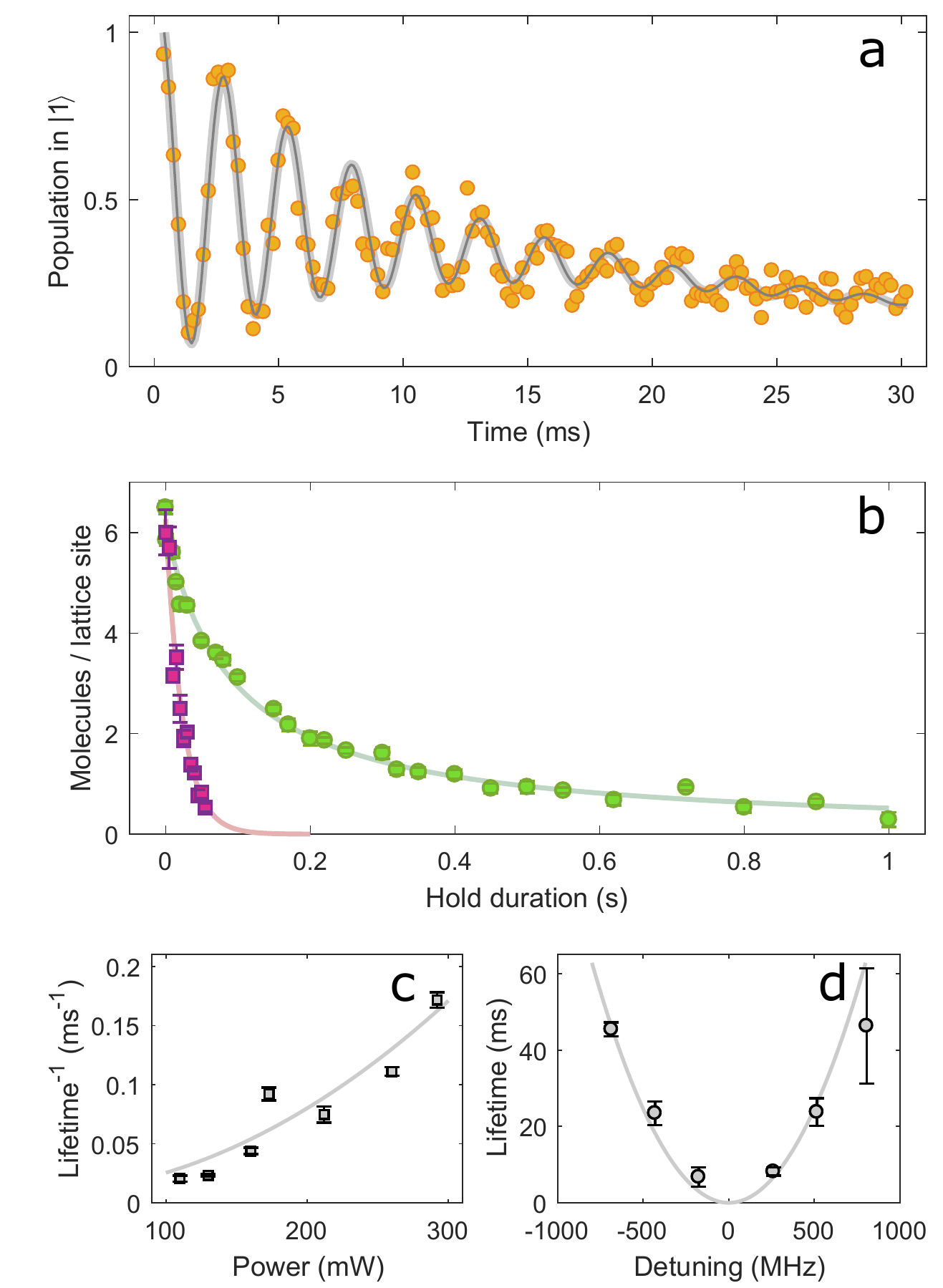}
\caption{Coherent control of molecular clock states.  a) Rabi oscillations exhibit two time constants:  an overall molecule lifetime of 30 ms and a Rabi coherence time of 10 ms.  Each point is an average of three experimental realizations, and the fitted Rabi rate is 388(1) Hz.  b) Molecular lifetimes for $|2\rangle$ are measured in a magic (red squares) and nonmagic (green circles) lattice.  In a nonmagic lattice, the loss is dominated by two-body collisions, the asymptotic molecule number per lattice site approaching 0.5 as expected.  In a magic lattice, the single-body lifetime is 24(2) ms from an exponential decay fit.  c) Inverse quadratic dependence of the lifetime on lattice power as indicated by a parabolic fit, suggesting that a limit is imposed by two-photon light scattering via photodissociation.  d) Quadratic dependence of the lifetime on lattice detuning from the magic frequency for $v'=22$ and $P_l=165$ mW.  Each point in (b) is an average of $\sim10$ shots and error bars correspond to the standard errors of the mean.  Each point in (c,d) is obtained from a fit to an exponential loss curve and error bars are the standard fit errors.}
\label{fig:Fig3}
\end{figure}
To demonstrate coherent quantum state control of the molecular clock, we induce Rabi oscillations between $|1\rangle$ and $|2\rangle$.  Figure \ref{fig:Fig3}(a) shows coherent oscillations persisting for $\sim30$ ms.  The measurement is fitted to an oscillation where both the particle number and the fringe contrast are decaying, as described in SI, and suggests that two time constants currently limit the clock $Q$.  The overall decay indicates a 30 ms lifetime of $|2\rangle$ in the magic lattice.  The coherence time that governs fringe contrast is 10 ms for this data set, and can be traced to technical causes such as short- and long-term instabilities in the probe laser frequency.

The near-resonant set point of the magic lattice is a concern for heating and loss of $|2\rangle$ by incoherent light scattering.  We directly measured the clock state lifetimes and found the $|2\rangle$ lifetime of $\sim200$ ms far from the polarizability resonance to be limited by bimolecular collisions with a two-body loss coefficient $\beta=2.1(6)\times10^{-10}$ cm$^3$/s.  The quoted error includes uncertainties in the molecule density and temperature as well as the standard error of the fit that is shown in Fig. \ref{fig:Fig3}(b) and described in SI.  As also shown in Fig. \ref{fig:Fig3}(b), in the magic lattice the $|2\rangle$ lifetime is limited by the lattice intensity, despite the fact that a significantly lower one-photon scattering rate is expected from \textit{ab initio} calculations \cite{MoszynskiSkomorowskiJCP12_Sr2Dynamics} of $1_u$ lifetimes and branching ratios.  Additional measurements revealed that the $|2\rangle$ lifetime drops as $\propto 1/P_l^2$ where $P_l$ is the lattice light power (Fig. \ref{fig:Fig3}(c)), suggesting two-photon scattering.  The two-photon process can couple $|2\rangle$ to a higher-energy continuum, dissociating the molecules.  For the current choice of lattice wavelength, the targeted \textit{gerade} continuum lies below the $^1S+{^1P}_1$ atomic threshold but above $^1S+{^1D}_2$.  While photodissociation of Sr$_2$ to lower-lying continua is well understood \cite{ZelevinskyMcDonaldNature16_Sr2PD}, calculating lattice-induced coupling to higher-lying continua is challenging due to the large number of possible molecular potential curves.  Our preliminary calculations yield two-photon scattering rates that are consistent with observations.  Furthermore, we expect that this scattering has a non-monotonic dependence on the continuum energy, and could be suppressed by a judicious choice of the intermediate $1_u$ state.  We note that the clock transition explored here spans nearly the entire molecular potential and thus requires compensation of a large lattice-induced light shift ($\sim20\%$ of the trap depth, comparable to the case of rotational qubits \cite{DeMilleKotochigovaPRA10_PolarMoleculesStateInsensTrapping,JinNeyenhuisPRL12_KRbAnisotropicPolarizability}).  Alternative scenarios based on more closely spaced molecular levels could require less polarizability tuning \cite{ZwierleinParkScience17_NaKNuclearSpinCoherence} and thus allow the lattice frequency to be further from a vibronic resonance, resulting in proportionally lower two-photon scattering rates.  A typical quadratic dependence of molecule lifetime on the lattice detuning from a polarizability resonance is presented in Fig. \ref{fig:Fig3}(d).

\begin{figure}
\includegraphics*[trim = 0in 4in 3.2in 3in, clip, width=3.375in]{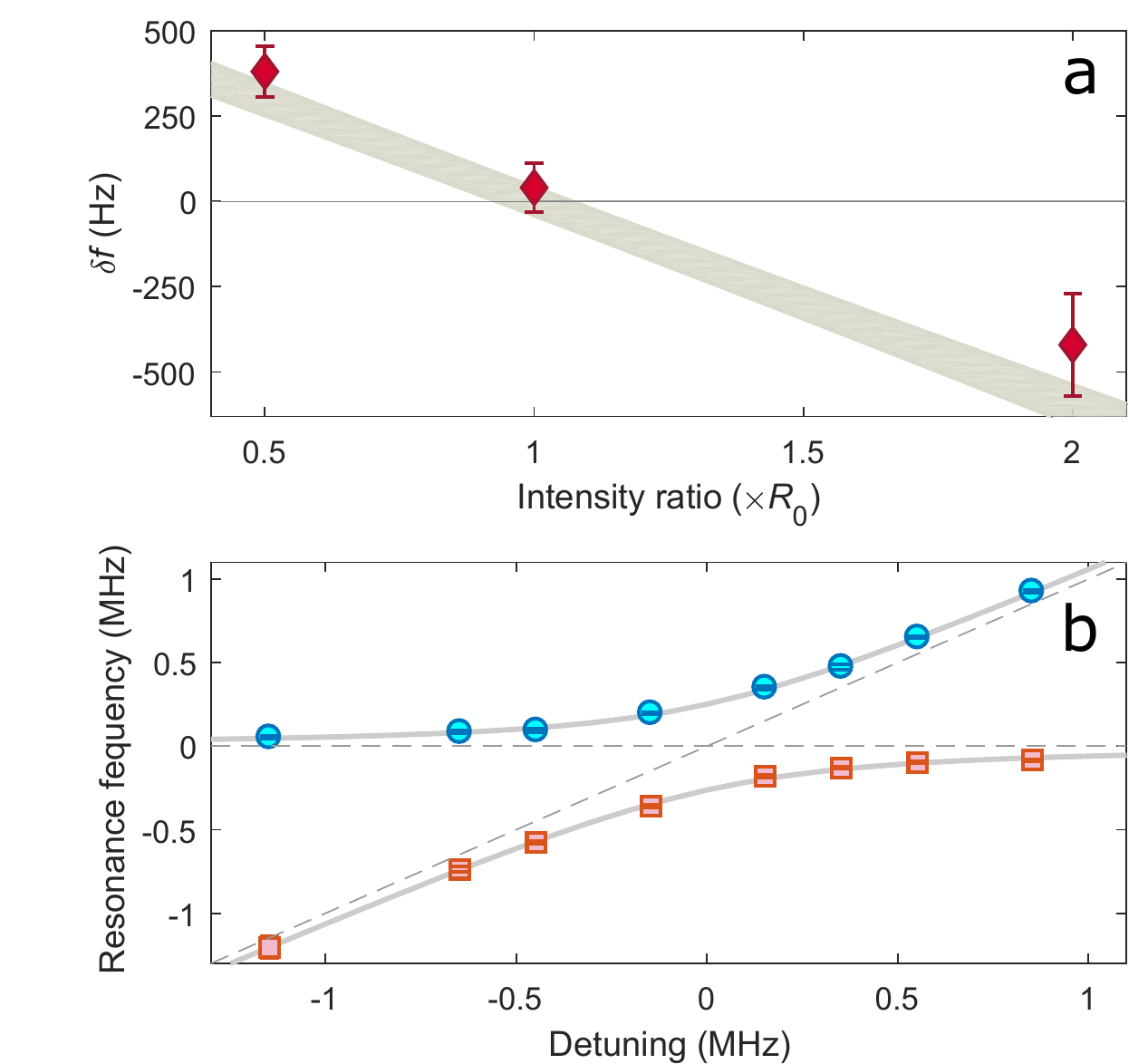}
\caption{Magic intensity ratio for a two-photon clock transition.  a) At the probe intensity ratio where the Rabi rates are equal, the light shift is consistent with zero.  Here the light shift (with $\Omega$ varied between 46(3) and 1240(90) Hz) is measured for several intensity ratios.  The error bars quantify the uncertainties arising from the reference laser frequency drift during data collection, and the shaded region shows the expected shift calculated from Rabi frequency measurements as in panel (b).  b) Autler-Townes molecular spectroscopy reveals the avoided crossing curves, and allows precise measurements of transition Rabi frequencies and thus determination of the magic intensity ratios.  Here, weak 689.4 nm light is used as probe, and the coupling is provided by 651.9 nm light with a resonant Rabi rate given by the minimum separation of the hyperbola branches ($\Omega_2=514(9)$ kHz).  The uncertainties are standard errors of the clock line centers when fitted to Lorentzian curves.}
\label{fig:Fig4}
\end{figure}
With $Q$ approaching $10^{12}$, systematic effects in the molecular clock can be studied with a high precision.
This is particularly important due to the basic differences between molecular and atomic lattice clocks, where the systematics were recently suppressed to the $\sim10^{-18}$ level \cite{YeCampbellScience17_3DClock,LudlowSchioppoNPhot17_DualYbClock,KatoriNemitzNPhot16_YbSrClockRatio}, including the two-color probe scheme.
Here we study the effects of the probe light on the clock transition in this configuration where both off-resonant Raman beams light-shift the clock states.  This shift can be strongly suppressed by balancing the intensity ratio of the two beams.  To the lowest order, which is easily resolved with the current clock precision, the differential light shift is nulled if the effective Rabi frequencies associated with the two probe beams are equal.  Figure \ref{fig:Fig4}(a) shows that the probe-beam power ratio corresponding to this balanced condition, $R_0$, results in a differential light shift consistent with zero.  For each Rabi frequency ratio, the probe intensities were varied by a factor of 27.  At ratios of $0.5R_0$ and $2R_0$, a substantial net clock shift of several hundred Hz is observed.  To implement the probe light shift cancellation scheme we used Autler-Townes spectroscopy to precisely measure single-photon Rabi rates in the molecular clock.  This method results in avoided-crossing curves described in SI and shown in Fig. \ref{fig:Fig4}(b), where the Rabi rate $\Omega$ is the distance between the hyperbola branches at the intermediate-state detuning $\Delta=0$.  The measurements indicate that our $1\%$ control of the probe intensity ratio should eliminate these light shifts at the $2\times10^{-15}$ level for the 25.1 THz clock transition.  Substantial improvement can be achieved with lower probe intensities at longer molecule-light coherence times, or with additional cancellation methods such as supplying probe light that is both red- and blue-detuned from the intermediate state.

The reported vibrational molecular lattice clock demonstrates that precision measurements and frequency metrology based on new types of quantum dynamics are possible when a high level of molecular quantum state control is attained.  The long coherence times observed here across a large vibrational energy gap are directly applicable to molecular qubits and quantum simulators \cite{YeMosesNPhys17_PolarMoleculeFrontiers,YeYanNature13_KRbLatticeSpinModel,ZwierleinParkScience17_NaKNuclearSpinCoherence} where the long duration of information storage and the relatively small strength of dipolar interactions require extended observations.  Molecular state-insensitive lattice traps based on narrow vibronic resonances, described here, present several challenges that are expected to be overcome, as well as some advantages.  The latter include a high density of polarizability resonances and thus the ease of locating many candidate resonances in a convenient wavelength range.  The singlet-triplet character of Sr$_2$ lattice-driven vibronic transitions results in resonances that are much narrower than their spacing and thus do not interfere, presenting a clean platform for trap shift compensation.  This general property is also found in KRb and other ultracold molecules of interest.  Further work on precision, accuracy, and systematic effects is necessary to realize the full potential of vibrational molecular lattice clocks.
The level of precision achieved here is directly applicable to new fundamental measurements \cite{UbachsSalumbidesPRD13_FifthForcesMolecPrecisMeasts,TakahashiBorkowskiPRA17_YbPABeyondBO,ZelevinskyPRL08,ChardonnetShelkovnikovPRL08_muStability,UbachsSalumbidesPRL11_H2MoleculeQED}.

\section{Acknowledgments}

We gratefully acknowledge the NSF grant PHY-1349725 and the ONR grant N00014-17-1-2246, as well as the Polish National Science Center Grant 2016/20/W/ST4/00314.

\appendix*

\section{Supplementary information}

\subsection{Details of the molecular clock experiment}
\label{sec:Experiment}

Strontium atoms are cooled on the strong 461 nm transition, transferred to a magneto-optical trap operating on the intercombination line at 689 nm, and loaded into a one-dimensional retro-reflected optical lattice with a wavelength that is continuously tunable between 907 and 926 nm.  Subsequently we photoassociate the atoms into excited-state $0_u^+(v'=-4,J'=1,M'=\pm1)$ Sr$_2$ molecules using a 29 W/cm$^2$  pulse of 2 ms duration, where the quantum numbers are defined as in the manuscript.  The excited molecules decay preferentially into the X$0_g^+(v=-1,J=\{0,2\})$ ground states \cite{ZelevinskyReinaudiPRL12_Sr2}.  We produce $\sim7000$ molecules in ($v=-1,J=0$), spread over $\sim700$ lattice sites.

At an optical power of 155 mW and wavelength of 922 nm, the lattice depth for Sr$_2$ molecules is $\nu_z^2M^2\lambda_l^4/h^2=800$ $E_r$, where $E_r = 1.33$ $h\,$kHz is the lattice recoil energy, $\lambda_l$ is the lattice wavelength, $M$ is the molecular mass, and $h$ is the Planck constant.  On each lattice site, the axial and radial confinement frequencies are $\nu_z=74$ kHz and $\nu_r=440$ Hz, respectively.  The temperature of the gas depends on the details of preparation and is in the 4-8 $\mu$K range for this study.  The axial frequency and temperature were extracted from subradiant molecular spectroscopy \cite{ZelevinskyMcDonaldPRL15_Sr2LatticeThermometry}.  The combination of a large lattice depth and relatively low temperature ensures that tunneling is strongly suppressed.

Spectroscopy and coherent manipulation of the clock states are performed with a two-photon anti-Stokes Raman process.  The Raman lasers are tuned 25 MHz above the $0_u^+(v'=-4,J'=1)$ intermediate state and copropagate with the lattice.  Their wavelengths are 689.4496 and 651.8623 nm, as measured by a wavemeter.  The vertical quantization axis is set by a small magnetic field, and the lattice and Raman beam polarizations point along this axis.  The lattice beam waist where the molecules are trapped is 30 $\mu$m, while the probe beam waists are 90 $\mu$m, providing uniform illumination over the typical cloud size of $<15$ $\mu$m.  After clock spectroscopy is performed, the molecules in the upper clock state $|1\rangle$ are photodissociated into slow Sr atoms using a 2 ms pulse with peak intensity 5.3 W/cm$^2$ tuned above the singly-excited $^1S{+^3P}_1$ continuum.  The atoms are counted in an absorption image using a 50 $\mu$s pulse of resonant 461 nm light.

The lattice light is derived from a tapered amplifier diode.  At a magic wavelength, resonant scattering due to amplified spontaneous emission can shorten the molecule lifetime.  We mitigate this effect by spectrally filtering the lattice light using a linear Fabry-Perot cavity with a finesse of 160 and a free spectral range of 2.6 GHz.

For the high-$Q$ clock spectrum in Fig. \ref{fig:Fig2}(b), the intensities of the 689.4 and 651.9 nm Raman lasers are 0.16 mW/cm$^2$ and 5.8 mW/cm$^2$, respectively, and the axial trap frequency is reduced to 59 kHz.  The Rabi oscillations in Fig. \ref{fig:Fig3}(a) were measured with probe light intensities that are larger by $\sim40\times$, or 6.8 mW/cm$^2$ and 224 mW/cm$^2$ for the two colors, respectively.

\subsection{Vibronic polarizability resonances}
\label{sec:PolarizabilityResonances}

\begin{figure}
\includegraphics*[trim = 0.5in 5.5in 2in 2in, clip, width=3.375in]{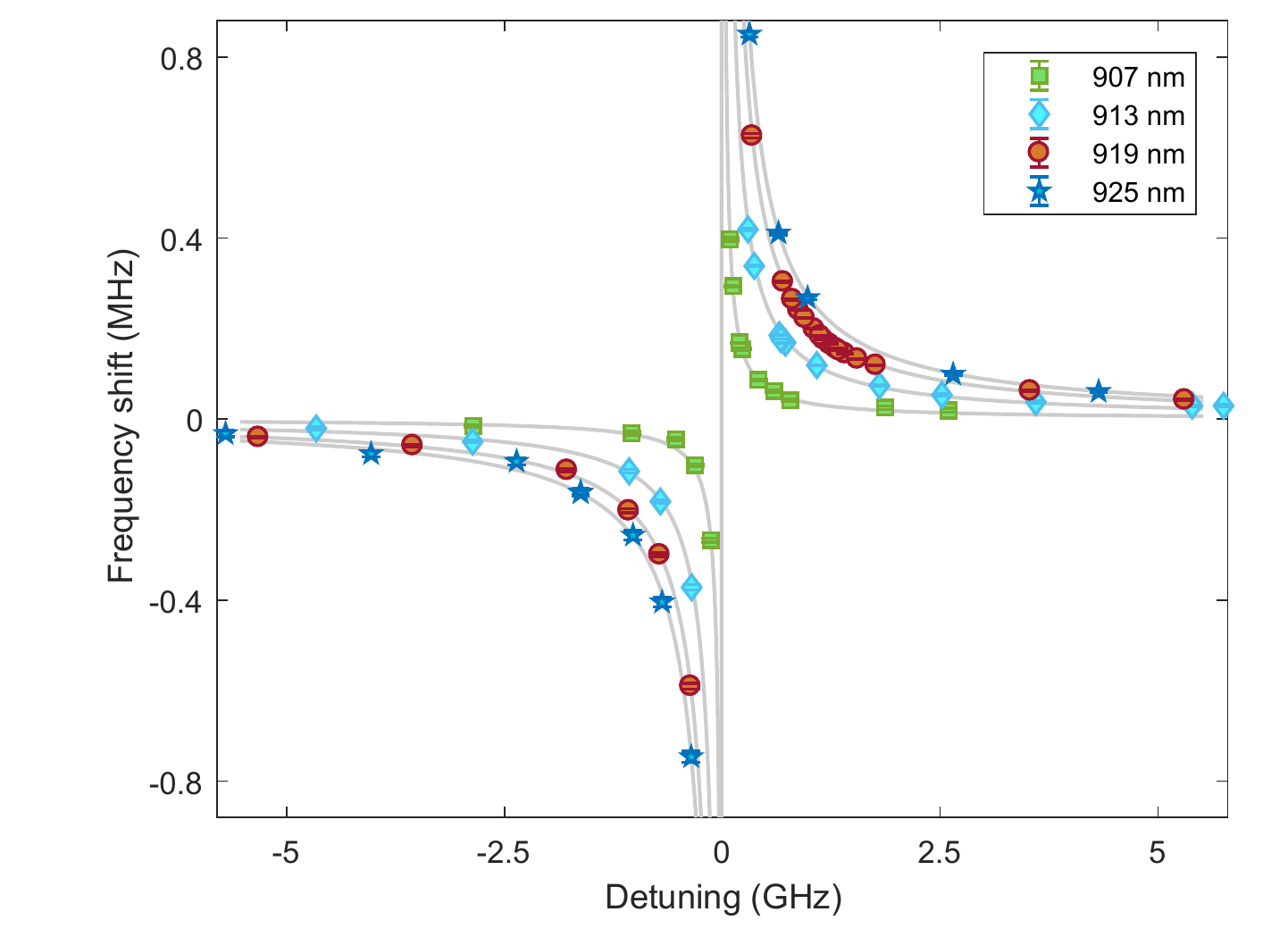}
\caption{Measured light shifts of clock state $|2\rangle$ versus lattice light detuning shows four polarizability resonances originating from coupling to $1_u(v'=\{19$-$22\})$ at 925-907 nm wavelengths.  The vertical axis refers to the detuning of the 689.4 nm probe laser, and the measurements were performed at lattice power of 200-230 mW.  The dispersive fitting function is described below.  The error bars correspond to the standard errors of Lorentzian fits to the clock spectra.}
\label{fig:Fig1S}
\end{figure}
In addition to the polarizability resonance of clock state $|2\rangle$ shown in Fig. \ref{fig:Fig2}(a), we have measured three additional resonances, as shown in Fig. \ref{fig:Fig1S}.  The lattice couples $|2\rangle$ to $1_u(v'=\{19$-$22\})$, where $v'=20$ was mostly used in this work.

\subsection{Functions used to describe the data}
\label{sec:Functions}

\subsubsection{State-insensitive trapping}
\label{sec:Fig2}

The lattice-induced light shift of $|2\rangle$ is described by
\be
\delta f = \frac{1}{4}\frac{\Delta\Omega_l^2}{\Omega_l^2/2+\Delta^2+\gamma^2/4} \approx \frac{\Omega_l^2}{4\Delta},
\ee
where $\Omega_l$ is the resonant lattice-induced Rabi frequency, $\Delta$ is the detuning from the nearest $1_u$ level and $\gamma$ is its natural width.  The function used to fit the data in Figs. \ref{fig:Fig2}(a) and \ref{fig:Fig1S} is
\be
y(x) = \frac{A}{x-x_0}+B.
\ee

The clock spectra that were collected to obtain the plot of linewidths $w$ in Fig. \ref{fig:Fig2}(b) were fitted to Lorentzian profiles
\be
y(x) = \frac{A}{1+[2(x-x_0)/w]^2}
\ee
after taking a natural logarithm of the Sr photofragment count derived from absorption images.
The same function was used to fit the 32 Hz wide clock line in Fig. \ref{fig:Fig2}(b), as well as the spectra measured for Fig. \ref{fig:Fig4}.

Two clock lineshapes in Fig. \ref{fig:Fig2}(b) exhibit broadening in a nonmagic lattice due to the residual temperature of the molecules.  They are described by the functional form $u^2e^{-u}$ \cite{ZelevinskyMcDonaldPRL15_Sr2LatticeThermometry}, where $u=\delta f / [k_BT(\sqrt{\alpha'/\alpha}-1)]$, and $\alpha$ and $\alpha'$ are the polarizabilities of $|2\rangle$ and $|1\rangle$.  The fitting function is
\begin{equation}
y(x) = B - A(x-x_0)^2e^{-B(x-x_0)}.
\end{equation}

\subsubsection{Coherent Rabi oscillations}
\label{sec:Fig3a}

The function used to fit the oscillations in Fig. \ref{fig:Fig3}(a) at Rabi frequency $\Omega$, and to define the lifetime and coherence time constants $\tau_L$ and $\tau_C$, is
\begin{equation}
y(x) = A e^{-x/\tau_L}\left[1+e^{-x/\tau_C}\cos(\Omega x+\phi)\right].
\end{equation}

\subsubsection{Molecule loss near a magic wavelength}
\label{sec:Fig3bInset}

The inverse molecular lifetime in a magic lattice trap as a function of the lattice light power is fitted to
\be
y(x) = Ax^2
\ee
in Fig. \ref{fig:Fig3}(c).  The same function is used to fit the molecule lifetime as a function of the lattice detuning from the magic wavelength in Fig. \ref{fig:Fig3}(d).

\subsubsection{Molecule loss due to collisions}
\label{sec:Fig3b}

The molecular lifetime curve in Fig. \ref{fig:Fig3}(b), measured away from a trap-induced resonance (green points), is fitted to a function describing combined one-body and two-body losses,
\be
N(t)=\frac{N_0e^{-\gamma t}}{1+\frac{N_0A}{\gamma}(1-e^{-\gamma t})},
\label{eq:TwoBodyFitCurve}
\ee
where $N_0$ is the initial molecule number per lattice site and $\gamma$ is the one-body exponential decay rate.
The molecule losses are dominated by the two-body loss coefficient
\be
\beta=\left(4\pi k_BT/M\right)^{3/2}\frac{2A}{\omega_z\omega_r^2}.
\label{eq:BetaDefined}
\ee
In Eq. (\ref{eq:BetaDefined}), $k_B$ is the Boltzmann constant, $\omega\equiv2\pi\nu$ and $T$ is the molecular temperature which, along with the trap frequencies, is an independently measured parameter.  At relatively short times, the fit is not sensitive to the exponential lifetime $1/\gamma$, and returns $1/\gamma=4(5)$ s.
The resulting coefficient $\beta=2.1(6)\times10^{-10}$ cm$^3$/s could be overestimated since it does not consider the $J=2$ molecules that are present but unobserved.

\subsubsection{Autler-Townes spectroscopy}
\label{sec:Fig4a}

Two branches of the Autler-Townes doublet hyperbola, plotted in Fig. \ref{fig:Fig4}(b) for varying probe detunings \cite{ZelevinskyMcGuyerNJP15_Sr2Spectroscopy}, are individually fitted to
\begin{equation}
y=\frac{1}{2}(x-x_0)\pm \frac{1}{2}\sqrt{(x-x_0)^2+\Omega_2^2} +B,
\end{equation}
where $\Omega_2$ is the resonant Rabi rate associated with the 651.9 nm probe.

\subsection{Natural lifetimes of the vibrational states}
\label{sec:Lifetimes}

The dominant decay mechanism of the vibrational clock states in the electronic ground state, X$0_g^+$ ($J=0$, $v$), is the electric quadrupole (E2) transition to the lower-lying vibrational states.  The E2 selection rules ensure that $\Delta J\leq2$.  Moreover, $J=0\rightarrow J=0$ is forbidden, and only even values of $J$ are allowed due to the bosonic statistics of $^{88}$Sr.  Therefore, the X$0_g^+$ ($J=0$, $v$) levels decay to X$0_g^+$ ($J=2$, $v-1, v-2, ...$).

\begin{figure}
\includegraphics*[trim = 1in 4in 1.5in 4in, clip, width=3.375in]{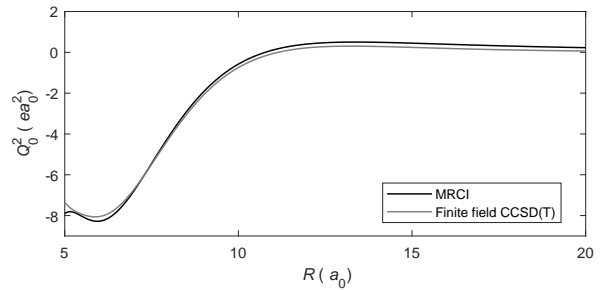}
\caption{The quadrupole moment of the $^{88}$Sr$_2$ ground state calculated using two distinct quantum chemistry methods.}
\label{fig:Fig2S}
\end{figure}
The natural lifetime $\tau$ is given by the spontaneous emission rate as $\tau = 1/\gamma$.  The spontaneous emission rate is obtained from the Einstein A coefficients,
\begin{align}
 \gamma_{J=0, v''} = \sum_{v'}A^{J=0, v''}_{J=2, v'}.
\end{align}
The Einstein A coefficients for the E2 transitions are \cite{SobelmanAtomicSpectraBook}
\begin{align}
\nonumber A^{J=0, v''}_{J=2, v'} = \frac{1}{15} \frac{1}{4 \pi \varepsilon_0} \frac{(E_{J=0, v''} - E_{J=2, v'})^5}{\hbar^6 c^5}\times \\ H^{J=0}_{J=2} |\langle \chi_{J=0, v''}(R) | Q^2_0(R) | \chi_{J=2, v'}(R)  \rangle|^2,
\end{align}
where $R$ is the internuclear distance, $E$ is the energy, $H^{J=0}_{J=2}$ is the Honl-London factor obtained by integrating the angular parts of the wave functions, and $\chi(R)$ are the rovibrational wave functions.  The quadrupole moment $Q^2_0(R)$ is plotted in Fig. \ref{fig:Fig2S}. It was calculated using two distinct quantum chemistry methods, multireference configuration interaction (MRCI) and finite field coupled cluster (CCSD(T)).  The resulting lifetimes differ at most by 10\%.  The finite field CCSD(T) calculations were ultimately used.

\begin{figure}
\includegraphics*[trim = 1in 4in 1.5in 4in, clip, width=3.375in]{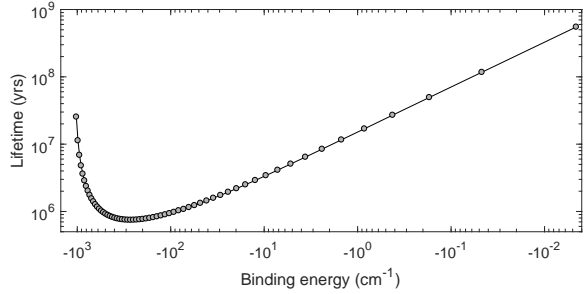}
\caption{Lifetimes of the X$0_g^+$ ($J=0$, $v$) states.}
\label{fig:Fig3S}
\end{figure}
The resulting lifetimes of the X$0_g^+$ ($J=0$, $v$) states are shown in Fig. \ref{fig:Fig3S}.  For $v=6$, the lifetime is 2.9 million years.  The shortest lifetime, for $v\sim25$, is 700,000 years.  The lowest vibrational states have longer lifetimes since they have few available decay channels.  The lifetimes are also longer for weakly bound states since the quadrupole moment vanishes for large interatomic separations.


\end{document}